\documentclass[journal]{IEEEtran}
\ifCLASSINFOpdf
\else
\fi
\usepackage{amsmath}
\usepackage{amssymb}
\usepackage{booktabs}
\usepackage{tabularx}
\usepackage{CJKutf8}
\usepackage{graphicx}
\begin{document}
%
% paper title
% Titles are generally capitalized except for words such as a, an, and, as,
% at, but, by, for, in, nor, of, on, or, the, to and up, which are usually
% not capitalized unless they are the first or last word of the title.
% Linebreaks \\ can be used within to get better formatting as desired.
% Do not put math or special symbols in the title.
\title{MetaAID 2.5: A Secure Framework for Developing Metaverse Applications via Large Language Models}
%
%
% author names and IEEE memberships
% note positions of commas and nonbreaking spaces ( ~ ) LaTeX will not break
% a structure at a ~ so this keeps an author's name from being broken across
% two lines.
% use \thanks{} to gain access to the first footnote area
% a separate \thanks must be used for each paragraph as LaTeX2e's \thanks
% was not built to handle multiple paragraphs
%

\author{Hongyin~Zhu

\thanks{H. Zhu (e-mail: hongyin\_zhu@163.com).}
% \thanks{P. Tiwari (e-mail: prayag.tiwari@ieee.org).}
% \thanks{H. Zhu is with the Inspur Electronic Information Industry Co., Ltd., Beijing 10010, China, also with the National Key Laboratory of High-end Server System, Jinan 250013, China. (e-mail: hongyin\_zhu@163.com).}% <-this % stops a space
% \thanks{P. Tiwari is with the School of Information Technology, Halmstad University, Sweden (prayag.tiwari@ieee.org).}% <-this % stops a space
% \thanks{Manuscript received April 19, 2005; revised August 26, 2015.}
}

% note the % following the last \IEEEmembership and also \thanks - 
% these prevent an unwanted space from occurring between the last author name
% and the end of the author line. i.e., if you had this:
% 
% \author{....lastname \thanks{...} \thanks{...} }
%                     ^------------^------------^----Do not want these spaces!
%
% a space would be appended to the last name and could cause every name on that
% line to be shifted left slightly. This is one of those "LaTeX things". For
% instance, "\textbf{A} \textbf{B}" will typeset as "A B" not "AB". To get
% "AB" then you have to do: "\textbf{A}\textbf{B}"
% \thanks is no different in this regard, so shield the last } of each \thanks
% that ends a line with a % and do not let a space in before the next \thanks.
% Spaces after \IEEEmembership other than the last one are OK (and needed) as
% you are supposed to have spaces between the names. For what it is worth,
% this is a minor point as most people would not even notice if the said evil
% space somehow managed to creep in.

% The paper headers
\markboth{Journal of IEEE}%
% \markboth{Journal of \LaTeX\ Class Files,~Vol.~14, No.~8, August~2015}%
{Shell \MakeLowercase{\textit{et al.}}: Bare Demo of IEEEtran.cls for IEEE Journals}
% The only time the second header will appear is for the odd numbered pages
% after the title page when using the twoside option.
% 
% *** Note that you probably will NOT want to include the author's ***
% *** name in the headers of peer review papers.                   ***
% You can use \ifCLASSOPTIONpeerreview for conditional compilation here if
% you desire.

% If you want to put a publisher's ID mark on the page you can do it like
% this:
%\IEEEpubid{0000--0000/00\$00.00~\copyright~2015 IEEE}
% Remember, if you use this you must call \IEEEpubidadjcol in the second
% column for its text to clear the IEEEpubid mark.

% use for special paper notices
%\IEEEspecialpapernotice{(Invited Paper)}

% make the title area
\maketitle

% As a general rule, do not put math, special symbols or citations
% in the abstract or keywords.
\begin{abstract}
Large language models (LLMs) are increasingly being used in Metaverse environments to generate dynamic and realistic content and to control the behavior of non-player characters (NPCs). However, the cybersecurity concerns associated with LLMs have become increasingly prominent. Previous research has primarily focused on patching system vulnerabilities to enhance cybersecurity, but these approaches are not well-suited to the Metaverse, where the virtual space is more complex, LLMs are vulnerable, and ethical user interaction is critical. Moreover, the scope of cybersecurity in the Metaverse is expected to expand significantly. This paper proposes a method for enhancing cybersecurity through the simulation of user interaction with LLMs. Our goal is to educate users and strengthen their defense capabilities through exposure to a comprehensive simulation system. This system includes extensive Metaverse cybersecurity Q\&A and attack simulation scenarios. By engaging with these, users will improve their ability to recognize and withstand risks. Additionally, to address the ethical implications of user input, we propose using LLMs as evaluators to assess user content across five dimensions. We further adapt the models through vocabulary expansion training to better understand personalized inputs and emoticons. We conduct experiments on multiple LLMs and find that our approach is effective.
\end{abstract}

% Note that keywords are not normally used for peerreview papers.

% For peer review papers, you can put extra information on the cover
% page as needed:
% \ifCLASSOPTIONpeerreview
% \begin{center} \bfseries EDICS Category: 3-BBND \end{center}
% \fi
%
% For peerreview papers, this IEEEtran command inserts a page break and
% creates the second title. It will be ignored for other modes.
\IEEEpeerreviewmaketitle

\section{Introduction}
% 近一年AI发展速快很快，特别是大语言模型为主的技术，引起了资本市场的关注。在元宇宙中实现可信鲁棒安全的智能系统，对于用户交互的安全性是必不可少的因素。其中本文提出了使用大语言模型对用户进行培训教育实现防止数据窃取的方法。本文来介绍如何在共享环境中实现网络安全。网络刺探预防策略，通过使用大语言模型生成许多测试数据，来提前预演和模拟，并且进行模型微调，从而解决这个问题。进行模拟和演练，实现更快、安全、创新的软件。通过测试用例的自动生成，从而实现可解释的安全防护策略。

% 先前的工作主要基于云计算进行大语言模型的托管，他们没有将模型部署到AI PC。最近的工作开始将大语言模型部署到端侧，进行模型压缩。然而他们没有解决AIPC的网络安全问题。英伟达通过数字孪生，建设虚拟的数据中心仿真模拟，从而实现了真实策略的落地。在元宇宙大语言模型、智能体落地过程中，我们也需要借助大模型提供虚拟仿进行多次实验。并且根据仿真的表现进行多次优化模型的性能，在这个过程中面临的问题是。本文通过虚拟仿真技术来解决网络安全问题的枚举。我们自己的模型不可避免的会遭受网络刺探和信息泄漏。本文提出通过大语言模型生成虚拟的网络刺探数据，从而进行模拟和演练的方法。此问题的难点在于如何将模拟演练的过程中提高用户的安全意识和防御能力。

% 任何的技术都存在漏洞，因此要提高用户抵御风险的能力。因此本文提出带有用户安全培训能力的大模型。以及带有用户安全管控的AIPC大模型。为了解决用户安全培训问题，我们提出了基于提示工程的网络安全问题生成和答案生成算法。为了解决对用户使用行为的管控，我们提出了输入预估算法，从而预估用户的输入是否符合安全规范。

% 在过去的一年里，人工智能（AI）技术发展迅速，特别是大语言模型（LLM）技术，引起了学术界和产业界的广泛关注。人工智能发展的目标之一是实现可信、鲁棒、安全的智能系统。通过利用大语言模型，有望在元宇宙中率先实现此类系统。然而，大语言模型除了面临传统的网络安全问题外，还不可避免地会遭受网络刺探和信息泄露。为了实现这一目标，保障用户与大语言模型交互的安全性至关重要。
In the past year, artificial intelligence (AI) technology has advanced rapidly, particularly large language model (LLM) technology, which has garnered significant attention from both academia and industry. One of the objectives of the development of artificial intelligence is to create trusted, robust, and secure intelligent systems. It is hoped that by leveraging LLMs, these systems will be among the first to be implemented in the Metaverse \cite{zhu2022metaaid}. Users in the Metaverse can create avatars, interact with NPCs, and meet strangers, which exposes them to certain risks. In addition to traditional cybersecurity concerns, LLMs are also susceptible to network surveillance and information leaks. To achieve this, it is crucial to ensure the security of user interactions with LLMs.

% 先前的工作提出X方法，旨在增强大模型对对抗攻击的鲁棒性。深信服和奇安信两家公司分别开发了专注于网络安全的大语言模型。然而，这些模型在应用上主要局限于系统的自主防御和知识问答。在元宇宙中，我们期望大语言模型不仅能模拟网络刺探行为，还能根据模拟结果为用户提供有价值的建议。
Previous work has adopted adversarial training methods \cite{jain2023baseline} to enhance the robustness of large models against adversarial attacks. Rezanejad et al. \cite{rezanejadnew} propose using LLM to detect SQL injection attacks \cite{lakhani2022detecting}. Some companies, e.g. Sangfor Technologies (Security GPT) and Qi-AnXin Technology (Q-GPT), have developed LLMs with a focus on cybersecurity. However, the applications of existing models are primarily limited to the autonomous defense of systems and knowledge question-answering. These methods are designed for traditional network environments and are not suitable for the Metaverse \cite{ramalingam2023gpt}, where user activities and the creation of digital works are prevalent. In the Metaverse, anyone can potentially act as a hacker \cite{holoyad2023ml}. Therefore, we need to ensure data rights and limit user behavior, which broadens the concept of cybersecurity. In the Metaverse, we expect large language models to simulate network snooping behavior and provide valuable suggestions to users based on the simulation results.

% 在元宇宙中保障用户交互的安全性是个非平凡的问题。首先，第一个挑战是提高用户的安全意识和防御能力。我们的解决方案是利用大语言模型来教育和培训用户，以防止数据泄露，从而实现可解释的安全防护策略。通过生成大量的模拟数据和网络探测预防策略，用户可以进行一系列的模拟演练。这些演练不仅包括测试用例的自动生成，还包括模拟的网络攻击程序，以便用户能够亲身体验攻击的机制和后果。通过这种方式，用户能够在XR元宇宙环境中更好地理解并应对潜在的安全风险，从而提高他们的自我保护能力。
Securing user interactions in the Metaverse is a non-trivial task. The first challenge is to enhance users' security awareness and defense skills. We propose to use large language models to educate and train users to prevent data breaches, and to implement security protection policies that are transparent and easy to understand. By developing a system that contains a vast cybersecurity Q\&A corpus and simulated attack codes, users can engage in a series of practice exercises. These simulations include not only theoretical knowledge but also practical attack code scenarios, allowing users to experience firsthand the mechanics and consequences of an attack. In this way, users can better understand and respond to potential security risks in the XR Metaverse environment, thereby enhancing their self-protection abilities.

% 面临的第二个挑战是减少用户的违规输入。这其中需要解决的问题包括用户输入的评价，同时要兼容元宇宙中的个性化输入和表情包。为此，我们提出了LLM自动评价算法，基于元宇宙的社会属性，通过采用LLM作为评估器对用户的输入从5个维度进行评价，解决用户输入的安全问题。在元宇宙中，用户的输入是个性化的，类型非常丰富的，包括小众语言和表情包，已有的模型无法适应这种场景。因此我们提出了词汇扩展训练方法，通过扩展词汇表并结合高效的参数微调，最大程度保留模型的知识，并能迁移到各种不同场景，满足个性化输入需求。评估用户输入是一个高频且功能单一的场景，为了减少算力消耗，我们采用小模型+词汇扩展训练的方法，实现高性能计算效果。
The second challenge is to address unethical user input. We need to develop methods to evaluate user input and ensure compatibility with personalized inputs and emoticons within the Metaverse. To tackle this, we propose an automatic evaluation algorithm based on the social characteristics of the Metaverse and use LLM as an evaluator. This algorithm assesses user inputs across 5 dimensions. In the Metaverse, user input is diverse and personalized, encompassing various languages and emoticons. Existing models are not equipped to handle such scenarios effectively. Therefore, we introduce a vocabulary expansion training (VET) method. By expanding the vocabulary and integrating it with parameter-efficient fine-tuning, we can preserve the model's knowledge to a great extent and adapt to different scenarios to meet personalized input needs. Evaluating user input is a frequent and specialized scenario. To minimize computational power consumption, we combine small-sized LLMs with the vocabulary expansion training method to achieve high-performance computing.

% 本文针对多个大模型进行了实验，实验结果证明了方法的有效性。本文的主要创新点如下：
We conduct experiments on various LLMs, and the experimental results demonstrate the effectiveness of the method. The main innovations of this paper are as follows:

% （1）本文提出利用大语言模型提升用户人机交互安全性的思想。

% % （2）为实现提高用户的安全意识和防御能力，本文提出利用大语言模型生成模拟演练数据的方法。

% % （3）为实现用户输入评价和监控，本文提出LLM自动评价算法和词汇扩展训练方法。我们构建了网站，并取得了较好的实验结果。

(1) This paper proposes the idea of using LLMs to enhance the cybersecurity of user interactions within Metaverse applications.

(2) To enhance users' awareness and skills in defending against cyber threats, we propose using LLMs to simulate real-world scenarios for educational purposes. This includes interactive Q\&A sessions about cybersecurity and the simulation of attack codes to help users understand how attacks work.

(3) To ensure that user input is ethical, we propose using LLM to automatically evaluate user input. Additionally, we propose a vocabulary expansion training method. We conduct experiments on various LLMs, and the results demonstrate the effectiveness of our method.

\section{Related work}
\subsection{Cybersecurity Assessment of Large Language Models}
% 他们系统性地评估了大型视觉语言模型（VLMs）在面对视觉输入时生成响应的鲁棒性。他们提出了一种针对VLMs的对抗性攻击评估方法，即使在黑盒访问条件下也能有效进行。他们发现通过操纵视觉输入，可以自动化地欺骗VLMs生成目标响应。
Zhao et al. \cite{zhao2023evaluating} conducted a systematic evaluation of the robustness of large visual language models (VLMs) in generating responses to visual input. They developed an adversarial attack evaluation method that is effective even when the models are only accessible through their black-box interfaces. The researchers discovered that by manipulating the visual input, they could automatically trick VLMs into generating specific target responses. 
% 他们提出PRIV QA多模态基准，用于评估模型在保护个人信息的指令下的表现。他们引入自我调节技术，提高模型遵循访问控制指令的能力。他们发现模型在对抗性输入下的脆弱性，以及保护不同群体信息时的偏见。
Chen et al. \cite{chen2023can} introduce the PRIV QA multimodal benchmark to assess how well models can handle directives related to personal information protection. They propose self-regulatory methods to enhance the model's compliance with access control instructions. The study also identifies vulnerabilities in models to adversarial inputs and biases in protecting information from different social groups.

% 他们探讨了嵌入模型在映射原始数据到低维向量时，如何保留重要语义信息以及可能泄露的敏感信息。他们提出了三种攻击类型来系统研究嵌入模型可能泄露的信息：嵌入反演、敏感属性泄露和成员关系信息泄露。揭示了嵌入模型在处理敏感数据时的潜在隐私风险，并提出了相应的防御机制。
Song et al. \cite{song2020information} explored how embedding models preserve important semantic information and may leak sensitive information when mapping raw data to low-dimensional vectors. They proposed three attack types to systematically study the information that embedding models might leak: embedding inversion, sensitive attribute leakage, and membership information leakage. Their work revealed the potential privacy risks of embedding models when processing sensitive data and proposed corresponding defense mechanisms. 
% 他们介绍了一种名为ProPILE的新工具，用于检测大型语言模型（LLMs）中个人身份信息（PII）的泄露风险。ProPILE允许数据主体通过设计特定提示来评估LLM服务中潜在的PII泄露。
Kim et al. \cite{kim2023propile} introduce a new tool called ProPILE for detecting the risk of personally identifiable information (PII) leakage in LLMs. ProPILE allows data subjects to assess potential PII leaks in LLM services by designing specific prompts.
% 本文提出了一种名为"训练数据泄露报告"的方法，用于研究在强大的现实威胁模型下，当模型被提示相关上下文时可能泄露的训练数据中的用户内容。我们引入了一组特征，以帮助评估从用户级别隐私的角度来看可泄露的内容。我们还提出了两种度量标准，可以量化在相同数据上训练的不同模型的隐私泄漏情况。通过在电子邮件和论坛对话等真实世界数据集上进行大量数值研究，我们展示了我们的方法如何产生泄露报告，并进一步说明了如何利用提出的指标来研究缓解措施（如差异隐私培训和API加固）的效果。我们相信，可以将我们的框架纳入语言模型的训练平台，以帮助从隐私和实用性的角度评估模型。
Inan et al. \cite{inan2021privacy} introduce a set of features to help evaluate what can be leaked from a user-level privacy perspective. They also propose two metrics that can quantify the privacy leakage of different models trained on the same data.

% 他们探讨了语言模型在处理自然语言时面临的隐私挑战，指出现有数据保护技术不足以保障语言模型的隐私性。作者提出，应使用明确公开可用的数据来训练语言模型，以确保真正的隐私保护。
Brown et al. \cite{brown2022does} discuss the privacy challenges faced by language models when processing natural language. They argue that existing data protection technologies are not sufficient to ensure the privacy of language models. They propose that language models should be trained using only explicitly publicly available data to provide true privacy protection. 
% 他们探讨了语言模型（LM）在快速发展中所带来的隐私风险，并提出了一种新的技术调查框架，用于评估和比较不同类型的隐私攻击和防御策略。他们提出了一个全面的隐私攻击和防御策略分类框架，有助于研究者更好地理解和评估风险。
Smith et al. \cite{smith2023identifying} explore the privacy risks posed by the rapid development of language models (LMs) and propose a new technical investigation framework for evaluating and comparing different types of privacy attacks and defense strategies. They propose a comprehensive privacy attack and defense strategy classification framework to help researchers better understand and assess risks. 
% 他们详细探讨了已知的和预期的风险，并提出了六个主要风险领域：歧视与毒性、信息泄露、提供错误信息、恶意使用、人机交互危害以及自动化、访问和环境危害。论文分析了风险的来源，并提出了一系列风险缓解策略，包括技术解决方案、公共政策干预和社会参与。
Weidinger et al. \cite{weidinger2021ethical} provide a detailed examination of known and anticipated risks, identifying 6 primary risk categories: discrimination and toxicity, information leakage, dissemination of misinformation, malicious use, human-computer interaction hazards, and automation, access, and environmental hazards. 
% 他们提出了一种评估语言模型用户隐私泄露的方法，引入了量化用户数据泄露的指标。他们展示了如何利用这些指标来比较不同模型的隐私性。通过实验研究了RNN和Transformer模型的隐私泄露情况。探讨了使用差分隐私和API加固等技术来减少泄露。
Inan et al. \cite{inan2021training} proposed a method to evaluate the privacy leakage of language models and introduced indicators to quantify the leakage of user data. The study experimentally analyzed the privacy leakage of RNN and transformer models. Techniques such as differential privacy and API hardening were explored as potential methods to reduce such leakage. 
% 他们提出CV ALUES，首个评估中国大型语言模型（LLM）与人类价值观对齐情况的基准测试。包含对抗性安全提示和诱导性责任提示，覆盖广泛场景和领域。采用人类评价和自动评价相结合的方法，提供全面的价值评估。发现中国LLM在安全方面表现良好，但在责任方面有待改进。公开基准测试和代码，促进中国LLM价值观对齐研究。
Xu et al. \cite{xu2023cvalues} proposed CV ALUES, a benchmark designed to assess the alignment of Chinese LLMs with human values. The dataset includes confrontational safety tips and inductive responsibility tips, spanning various scenarios and fields. They found that Chinese LLMs demonstrated strong performance in terms of safety but required further improvement regarding responsibility. 
% 他们揭示了大型语言模型（LLMs）在下游应用中的安全风险，特别是在用户自定义微调时。通过实验发现，即使使用少量恶意设计的训练示例，也能轻易绕过模型的安全机制。提出潜在的安全性增强措施，并强调需要进一步研究来确保微调过程的安全性。
Qi et al. \cite{qi2023fine} highlight the security vulnerabilities of LLMs in various applications, particularly when fine-tuned by users. Their experiments demonstrate that with a minimal number of maliciously crafted training examples, the security mechanisms of these models can be compromised. 
% 他们提出了一个新指标来量化大型预训练语言模型中的潜在社会偏见和毒性内容。通过对24个常用模型进行实证分析，论文发现不同模型在代表性危害方面存在差异，并探讨了模型架构与危害之间的关系。创新点包括：1. 提出了一个基于语言建模的指标来评估代表性危害；2. 分析了24个模型的代表性危害，并与其他指标进行了比较；3. 发现模型深度与宽度对代表性危害有不同影响。
Hosseini et al. \cite{hosseini2023empirical} propose a new metric to quantify potential social bias and toxic content in large pre-trained language models. Through an empirical analysis of 24 commonly used models, they found that different models have differences in representative hazards, and explored the relationship between model architecture and hazards. 
% 他们提出了一个名为SafetyBench的全面的安全性评估基准，用于评价大型语言模型（LLMs）的安全性。该基准包含11,435个多选题，涉及7个不同的安全问题类别，并包括中文和英文数据。通过在零一和有限数据集上的实验，发现GPT-4在安全性评估中表现出显著优势。目前LLMs的安全性仍有很大提升空间。
Zhang et al. \cite{zhang2023safetybench} proposed a comprehensive safety assessment benchmark called SafetyBench for evaluating the safety of LLMs. The benchmark contains 11,435 multiple-choice questions covering 7 different security issue categories and includes Chinese and English data. Through experiments on zero-one and limited data sets, it is found that GPT-4 shows significant advantages in security evaluation.
% 他们系统地分析了大型语言模型（LLMs）的安全性和信任问题。 探讨了验证与验证（V&V）技术在确保LLMs安全性和信任性方面的应用。 提出了四种互补的V&V技术：虚假证明与评估、验证、运行时监控和法规与道德使用。 总结了LLMs的已知脆弱性和局限性，并讨论了如何通过V&V技术对其进行评估。 强调了在LLMs的生命周期中集成V&V技术的重要性，以确保其与安全性和信任性要求的符合性。
Huang et al. \cite{huang2023survey} conduct a comprehensive analysis of the security and trust challenges posed by LLMs. They explore the application of Verification and Validation (V\&V) techniques to ensure the security and trustworthiness of LLMs. 4 V\&V strategies are proposed: False Attestation and Assessment, Verification, Runtime Monitoring, and Regulatory and Ethical Use. These techniques are complementary and work together to address the various security and trust issues associated with LLMs.
% 他们全面概述了大型语言模型（LLM）的研究进展，包括预训练模型的细节、关键设计和开发方面以及性能差异等。文章还探讨了LLM在机器人技术、多模态LLM、增强型LLM、数据集和评估方面的应用。这篇综述性文章旨在为研究者提供一个快速全面的参考，以便更好地理解和利用这一技术。
Naveed et al. \cite{naveed2023comprehensive} provide a comprehensive overview of the research progress of LLMs. They discuss the details of pre-trained models, key design and development aspects, and performance differences. They also explore LLM applications in robotics, multimodal LLM, enhanced LLM, data sets, and evaluation.

\subsection{Attack and Defense of Large Language Models}
% 大模型的网络安全问题
% 他们探讨了大型语言模型（LLM）在安全与隐私领域的应用，并对其积极影响、潜在风险和内在脆弱性进行了分析。创新点在于系统且全面地总结了LLM在安全与隐私中的角色，并发现了LLM在增强代码和数据安全方面的显著优势。研究还分析了模型参数提取攻击研究不足的问题，并提出了安全指令调优作为新的防御机制。
Yao et al. \cite{yao2023survey} explore the application of LLMs in the field of security and privacy. They analyze the positive impacts, potential risks, and inherent vulnerabilities \cite{pearce2023examining} of LLMs in this context. They systematically and comprehensively summarize the role of LLMs in enhancing security and privacy, particularly in code and data security. 
% 他们提出了一种简单而有效的攻击方法，能够导致对齐的大型语言模型生成不适当的内容。他们使用贪婪搜索和梯度优化技术自动生成对抗性后缀，无需手动工程。他们开发的对抗性后缀具有良好的可迁移性，能在多个公开的大型语言模型上触发不适当的行为。
Zou et al. \cite{zouuniversal} propose a simple yet effective attack that causes aligned LLMs to generate inappropriate content. They use greedy search and gradient optimization techniques to automatically generate adversarial suffixes without manual engineering.
% 他们提出了一种新的方法，利用自然语言处理（NLP）技术，特别是大型语言模型（LLM），来检测和防止SQL注入攻击。该方法能够显著降低数据库的脆弱性，并有效抵御SQL注入黑客攻击。研究还展示了如何将LLM集成到Web应用防火墙（WAF）中，这是一种先进的防御技术。
Rezanejad et al. \cite{rezanejadnew} propose a new approach that utilizes LLMs to detect and prevent SQL injection attacks \cite{lodha2022sql}. This method can significantly reduce the vulnerability of the database and effectively resist SQL injection hacker attacks. This research also shows how LLM can be integrated into a web application firewall (WAF), an advanced defense technology. 
% They研究了大型语言模型（LLMs）中基于上下文的学习的安全性，特别是针对“演示攻击”的防御。作者提出了一种名为“advICL”的攻击方法，用于评估和增强基于上下文的学习的鲁棒性。实验表明，随着演示数量的增加，模型的鲁棒性会降低。此外，作者还提出了“Transferable-advICL”，这是一种可迁移的攻击方法，能够攻击未见的测试输入示例。
Wang et al. \cite{wang2023adversarial} study the security of context-based learning in LLMs, focusing on defense against "presentation attacks". They propose an attack method called "advICL" to assess and improve the robustness of context-based learning. 
% 他们区分了预训练语言模型中的记忆和关联能力。他们发现模型在记忆个人信息方面存在漏洞，但关联能力较弱，因此特定个人信息的泄露风险较低。他们提出防御技术来减轻潜在威胁，并提供新的视角来理解预训练语言模型的隐私风险，以促进模型安全性。
Huang et al. \cite{huang2022large} distinguish between memory and association capabilities in pre-trained language models. They found that the model had holes in its ability to remember personal information, but had weak correlation capabilities, so the risk of leaking specific personal information was low. They propose defense techniques to mitigate potential threats and provide new perspectives to understand the privacy risks \cite{kshetri2023cybercrime} of pre-trained language models to promote model security.

% 他们提出ReNeLLM框架，利用LLM自身生成高效、隐蔽的“越狱”提示。通过广泛的实验验证，ReNeLLM在攻击成功率、时间成本上显著优于基线方法。他们揭示了现有LLM防御方法的不足，并提供了详细的分析讨论。
Ding et al. \cite{ding2023wolf} proposed the ReNeLLM framework, which uses LLM itself to generate efficient and covert "jailbreak" prompts. 
% 他们提出了一种名为DeepInception的轻量级方法，用于破解大型语言模型（LLMs）的安全限制。该方法利用了LLMs的拟人化能力，通过构建新颖的场景来诱导模型违反安全规则。揭示了LLMs在自我失去控制方面的弱点。
Li et al. \cite{li2023deepinception} proposed a lightweight method called DeepInception for breaking the security constraints of LLMs. This method takes advantage of the anthropomorphic ability of LLMs to induce the model to violate safety rules by constructing novel scenarios. 
% GPTF UZZER 是一种自动化的黑盒式 jailbreak 测试框架，用于评估大型语言模型（LLMs）的安全性和可靠性。该框架能够生成针对 LLMs 的有害内容模板，并对其安全性进行测试。其自动化生成 jailbreak 模板的过程。
GPTF UZZER \cite{yu2023gptfuzzer} is an automated black-box jailbreak testing framework for evaluating the security and reliability of LLMs. This framework can generate harmful content templates targeting LLMs and test their security. 
% They提出了一种名为Prompt Automatic Iterative Refinement (PAIR) 的算法，用于生成能够绕过大型语言模型安全限制的提示。PAIR通过一个攻击者模型来迭代地优化这些提示，从而实现高效的“监狱逃脱”。PAIR在较少的查询次数内就能成功地绕过多个不同语言模型的安全机制，包括GPT-3.5/4、Vicuna和PalM-2。论文的创新点在于提出了一种高效的算法来识别大型语言模型中的安全漏洞，并生成能够绕过这些漏洞的提示。
Chao et al. \cite{chao2023jailbreaking} proposed an algorithm called Prompt Automatic Iterative Refinement (PAIR) for generating prompts that can bypass the security restrictions of LLMs. PAIR iteratively optimizes these hints through an attacker model to achieve an efficient "jail escape". PAIR can successfully bypass the security mechanisms of multiple different language models, including GPT-3.5/4, Vicuna and Palm-2, within a small number of queries. 

% 他们系统地分析了大型语言模型（LLM）的安全威胁模型和防御策略。他们评估了多种防御策略（包括检测、输入预处理和对抗训练）对LLM的实用性和有效性。他们探讨了LLM安全与计算机视觉领域的差异，并分析了防御策略在不同攻击场景下的有效性。他们提出了一种新的评估框架，用于比较LLM防御策略在白盒和灰盒攻击中的表现。
Jain et al. \cite{jain2023baseline} systematically analyzed the security threat models and defense strategies of LLMs. They evaluated the practicality and effectiveness of multiple defense strategies (including detection, input preprocessing, and adversarial training) against LLM. They explored the differences between LLM security and the field of computer vision and analyzed the effectiveness of defense strategies in different attack scenarios. 
% They提出了一种利用大型语言模型（LLMs）的对抗性后缀来欺骗模型生成危险响应的攻击方法，并提出了一种基于困惑度（perplexity）和token序列长度的分类器来检测这种攻击。创新点在于：1. 评估了困惑度在识别对抗性后缀攻击中的作用；2. 提出了一种结合困惑度和token序列长度的分类器，以减少误报并准确识别大多数对抗性攻击；3. 发现该方法能有效检测机器生成的对抗性后缀攻击，但对人类手工制作的攻击无效。
Alon et al. \cite{alon2023detecting} proposed an attack method that exploits adversarial suffixes on LLMs to trick the model into generating dangerous responses, and proposed a classifier based on perplexity and token sequence length to detect this attack. 
% 他们提出新的安全任务：受控代码生成，用于评估和增强大型语言模型（LLM）的安全性。他们开发了一种名为SVEN的学习方法，通过使用连续向量指导程序生成，无需修改LLM权重。他们创建了一个高质量的训练数据集，用于受控代码生成任务，同时对其他任务也有价值。
He et al. \cite{he2023large} propose a new security task: controlled code generation for evaluating and enhancing the security of LLMs. They developed a learning method called SVEN by using continuous vectors to guide program generation without modifying the LLM weights.

% They提出了一种新的方法来分析自然语言模型更新时泄露的信息，并提出了两个新的指标来量化这种泄露。通过比较两个模型，可以提取出训练数据中的敏感信息，即使这些信息只占原始数据的一小部分。论文还发现，在模型更新时添加或删除非敏感数据并不能有效减少泄露的风险。此外，使用差分隐私技术可以减少泄露，但会降低模型的实用性。
Zanella et al. \cite{zanella2020analyzing} propose a new method to analyze the information leaked when natural language models are updated, and propose two new metrics to quantify this leakage. By comparing the two models, sensitive information in the training data can be extracted, even if it only accounts for a small portion of the original data. 
% They提出了一种保护用户隐私的方法，通过检测和警告聊天机器人中可能泄露个人信息的可疑句子。论文提出了两种新的约束对齐模型，在PERSONA-LEAKAGE数据集上进行了评估，并分析了个性化聊天系统的行为。论文还构建了一个用于评估的新数据集PERSONA-LEAKAGE。
Xu et al. \cite{xu2020personal} propose a method to protect user privacy by detecting and warning suspicious sentences in chatbots that may reveal personal information. They propose two new constraint alignment models, evaluate them on the PERSONA-LEAKAGE data set, and analyze the behavior of the personalized chat system. 
% They提出了AttaQ数据集，用于评估大型语言模型对恶意或不当提问的反应。研究分析了不同模型的响应，并开发了自动方法来识别和定义模型中的脆弱区域。
Kour et al. \cite{kour2023unveiling} proposed the AttaQ dataset for evaluating the response of LLMs to malicious or inappropriate questions. They analyzed the responses of different models and developed automated methods to identify and define vulnerable areas in the model.
% They提出了一种名为MART的多轮自动红队训练框架，用于提高大型语言模型（LLMs）的安全性。该框架通过让一个对抗性的LLM与目标LLM交互，迭代生成更具挑战性的攻击性提示，同时对目标LLM进行安全对齐的微调。创新点在于：1. 引入自动化的红队训练，提高了效率；2. 通过多轮迭代，显著降低了安全风险；3. 保持了模型在非攻击性提示上的表现，即模型在遵循指令方面仍然表现良好。实验结果表明，MART能够有效降低违反安全准则的响应率，同时不损害模型的有用性。
Ge et al. \cite{ge2023mart} propose a multi-round automatic red team training framework called MART for improving the security of LLMs. This framework iteratively generates more challenging aggressive cues by letting an adversarial LLM interact with the target LLM while fine-tuning the target LLM for security alignment. 

\section{Preliminaries}
\subsection{Pre-trained Language Model}
% 大模型预训练与微调技术。
% 预训练技术是一种在大量文本数据上训练神经网络模型的方法，旨在使模型学习并理解语言的各种模式和关联。通过这种方式，预训练模型能够执行回答问题、生成文本或其他与语言处理相关的任务。这些模型通常基于Transformer架构，如BERT、GPT等，它们通过大规模的数据集进行训练，从而获得了广泛的语言理解和生成能力。
Pre-training technology involves training neural network models on vast amounts of text data to enable them to learn and understand various patterns and associations within language. The goal is to develop models that can perform tasks such as answering questions, generating text, and other language-processing tasks. These models are typically based on the Transformer architecture, such as BERT \cite{kenton2019bert} and GPT \cite{radford2018improving}, and are trained using a large corpus to acquire extensive language understanding and generation capabilities.

% 微调技术是一种在预训练语言模型基础上，针对特定下游任务进行进一步训练的方法。这种方法利用预训练模型所拥有的广泛知识和语言理解能力，通过在特定任务的数据集上进行精细调整（fine-tuning），使得模型能够更好地适应和解决特定问题。微调过程中，模型的参数会根据任务的具体需求进行调整，从而提高模型在特定应用场景中的表现。
Fine-tuning technology is a method of further training for specific downstream tasks, building upon a pre-trained language model. This approach leverages the extensive knowledge and language understanding capabilities of pre-trained models to better adapt and solve specific problems by fine-tuning task-specific data sets. During the fine-tuning process, the parameters of the model are adjusted according to the specific needs of the task, thereby improving the performance of the model in specific application scenarios.

% 预训练和微调技术的结合，使得大型语言模型能够广泛应用于各种自然语言处理的任务中，例如机器翻译、问答系统、文本生成和语言模型等。通过预训练，模型获得了广泛的通用知识，而微调则使得模型能够针对具体应用进行优化，从而提高了模型的准确性和适用性。
The combination of pre-training and fine-tuning techniques enables large-scale language models to be widely used in various natural language processing tasks \cite{zhu2022financial,zhu2023fqp}, such as machine translation, question-answering systems \cite{zhu2021collaborative}, text generation, etc. Through pre-training, the model acquires extensive general knowledge. Fine-tuning further optimizes the model for specific applications, thereby improving the accuracy and applicability of the model.

% 多模态大模型。
% 多模态预训练技术更进一步，它不仅处理文本数据，还涉及图像、视频和声音等其他数据模式。通过大规模的多模态数据对齐过程，对模型进行预训练，使得模型能够理解和生成更加丰富和复杂的信息。例如，模型可以同时理解一张图像和一段相关的文本，并能够生成同时包含视觉信息和文本描述的输出。这种技术使得模型能够理解和生成涉及多种模态的信息，包括文本、图像、视频和声音。
Multimodal pre-training technology takes a step forward by handling not only text data but also other data types such as images, videos, and audio. Through a large-scale multi-modal data alignment process, the model is pre-trained, allowing it to understand and generate more comprehensive and complex information. For example, a model can simultaneously understand an image and its corresponding text and generate an output that includes both visual information and a textual description. This technology enables models to understand and generate information across multiple modalities, including text, images, video, and audio.

% 随着大模型多模态预训练技术的发展，人工智能系统正日益逼近人类处理信息的方式，即通过多种感官的综合来理解和认知世界。随着技术的不断进步，未来的多模态预训练模型将能够处理更为复杂和多样化的任务，为人们的生活带来更多的便利和惊喜。
As multi-modal pre-training technology advances, AI systems are becoming closer to the way humans process information—by integrating multiple senses to understand and recognize the world. With ongoing technological advancements, future multi-modal pre-training models will be capable of handling more complex and varied tasks, bringing additional convenience and surprise to people's lives.

\subsection{Large Language Model}
% 提示工程。
% 大语言模型通常指的是参数规模达到十亿级别的模型。提示学习技术是一种利用这些大规模模型（如GPT-3）来执行各种任务的方法。通过设计特定的模板来引导模型的输出，用户可以指示模型生成文本、回答问题、总结文章，甚至是进行创作。这种技术不需要对模型进行重新训练，而是通过调整输入提示的方式来让模型适应新的任务。提示学习的关键在于设计有效的提示方式，以便模型能够理解并执行预期的任务。此外，还可以通过训练模板的表征参数来优化其对特定任务的响应。
Large language models typically refer to models with parameter scales reaching 1 billion or more. Prompt learning techniques are a way of leveraging these LLMs (like GPT-3) to perform a wide range of tasks. By designing specific templates to guide the model's output, users can leverage the model to generate text, answer questions, summarize articles, and even create creative works. This technique does not require retraining the model, but instead adapts the model to new tasks by adjusting the input prompts. The key to prompt learning is to design effective prompts so that the model can understand and perform the desired tasks. Additionally, the representation parameters of the prompt template can be trained to optimize its response to a specific task.

% 指令微调
% 指令微调技术是一种通过向大型语言模型提供一系列问答形式的指令，来调整模型输出的技术。这种技术旨在提高模型在特定应用或领域中的性能，使其能够更好地理解和响应用户意图，从而提供更加精准和有用的信息。模型通过学习理解和执行一系列问答形式的指令，获得了一定的适应性，这些指令可以是自然语言命令，也可以是更具体的编程指令。
Instruction tuning is a method that enhances a model's responses by training it on a substantial amount of question-answer pairs, which serve as guidelines. This approach aims to elevate the model's performance within a specific application or domain, enabling it to interpret and respond to user intentions more effectively. This technique leads to more accurate and relevant responses. LLMs become more versatile by learning to understand and execute instructions, which can be interacted with in the form of natural language commands or programming instructions.

% 代码生成
% 代码生成技术利用大语言模型来自动生成代码，通过大规模的数据集进行训练，以学习如何生成有效的代码序列。在代码生成过程中，模型会根据用户输入的上下文或描述来生成对应的代码段。这种技术可以显著提高软件开发的效率，特别是在自动化生成重复性高或基于模板的工作方面。然而，由于代码的复杂性和多样性，目前的代码生成技术还无法完全取代人类程序员，尤其是在需要高度创造性或复杂决策的编码任务中。
Code generation technology leverages the power of LLMs to automatically generate code. These models are trained on extensive code datasets, which often include open-licensed code from various online sources like GitHub, GitLab, and StackOverflow. Through this training, the models learn to produce sequences of code that are both effective and coherent. During the code generation process, LLMs create corresponding code snippets based on the context or description provided by the user. This technology can significantly improve the efficiency of software development, particularly in automating the generation of highly repetitive or template-based work. However, due to the complexity and diversity of code, current code generation technology cannot completely replace human programmers, especially in coding tasks that require high creativity or complex decision-making.

% In-context learning
% In-context learning技术使得大型模型能够通过观察特定的例子或上下文来学习新任务，而不需要直接的监督或大量的训练数据。这种技术使得模型能够从少量示例中推断出复杂的模式和关联，从而在新的情境中应用这些知识。In-context learning对于那些难以获取大量训练数据或者需要快速适应新任务的领域特别有用。通过这种方式，模型可以更好地理解和响应它以前从未见过的情境，从而提高了模型的泛化能力和适应性。
In-context learning technology empowers LLMs to acquire new skills by examining specific examples or contexts, bypassing the requirement for direct supervision or extensive training data. This approach equips models with the capacity to deduce intricate patterns and correlations from a limited number of examples, enabling them to apply this knowledge to novel scenarios. In-context learning is particularly advantageous in domains where it is challenging to amass large volumes of training data or where the ability to swiftly adapt to new tasks is paramount. In this manner, the model can enhance its understanding and response to situations it has not previously encountered, thereby augmenting the model's generalization capabilities and adaptability.

% Adaption（LoRA）
% Adaption技术是一种用于训练大规模语言模型的技术，其中LoRA（Low-Rank Adaptation）通过引入低秩适应的方法来提高模型的适应性和可扩展性。LoRA技术允许模型在不改变原始模型结构的情况下，通过学习低秩矩阵对模型进行微调，从而适应特定的任务或数据集。这种技术可以显著减少训练所需的时间和计算资源，同时保持较高的模型性能。LoRA技术在自然语言处理领域中得到了广泛应用，特别是在那些需要快速适应新任务或数据集的场景中。
Adaptation technology is used to fine-tune LLMs efficiently in environments with limited computing power. Low-rank adaptation (LoRA) is a technique that improves the adaptability and scalability of LLMs. The LoRA technology enables a model to adapt to a particular task or dataset by learning a low-rank matrix without modifying the original structure and without updating the majority of the model's parameters. This technique can significantly reduce the time and computing resources required for training while maintaining high model performance. LoRA technology is widely used in the field of domain adaptation, especially in scenarios that require rapid adaptation to new tasks or datasets.

\subsection{Reinforcement Learning from Human Feedback}
% RLHF
% RLHF技术是一种基于强化学习（Reinforcement Learning）和人类反馈（Human Feedback）的机器学习技术，用于改善和优化大型语言模型（如GPT-3）的行为和输出。该技术通过大规模的数据集进行训练，使模型能够理解和生成人类语言，并在与人类的交互中不断学习和适应。RLHF技术主要包括三个阶段：
Reinforcement learning from human feedback (RLHF) is designed to improve and refine the behavior and responses of LLMs. As the model interacts with humans, it continues to learn and adapt. Unlike previous methods, RLHF technology involves 3 special techniques. 
% 奖励模型训练（Reward Model Training）：通过收集大量的数据来训练一个奖励模型，该模型能够对语言模型的输出进行评分，以反映人类对输出的偏好。
(1) Reward model training: By collecting massive labeled samples to train a reward model, the model can score the output of the LLM to reflect human preferences for the output. 
% 强化学习训练（Reinforcement Learning Training）：使用奖励模型作为反馈机制，通过强化学习算法来优化语言模型。模型通过试错学习，不断调整参数，以产生更符合人类偏好的输出。
(2) Reinforcement learning training: By using the reward model as a feedback mechanism to optimize itself through reinforcement learning algorithms. It learns through trial and error, constantly adjusting its parameters to produce output that is more consistent with human preferences. 
% 人类反馈循环（Human Feedback Loop）：在训练过程中，人类反馈被用来微调模型的行为。模型会生成多个可能的回答，人类评估者会根据具体情况对这些回答进行排序或评分。这些反馈被用于更新模型的参数，从而引导模型产生更符合人类预期的结果。
(3) Human feedback loop: During the training process, human feedback is used to refine the model's behavior. The model generates several potential responses, which human evaluators rate or score based on the context. This feedback is then used to adjust the model's parameters, guiding the model to produce results that are more in line with human expectations. 
% 通过上述过程，大模型RLHF技术能够提高语言模型的质量、多样性和相关性，使其在回答问题、生成文本和执行其他语言任务时更加准确和自然。
By employing the above methodology, RLHF technology can enhance the quality, diversity, and relevance of LLMs. This leads to more accurate and natural responses when the models are answering questions, generating text, or performing other language-related tasks. 

% 对齐、评估、幻觉性
% 大模型对齐、评估和幻觉性是三个关键概念，它们对于确保大型语言模型（如GPT-3）安全、可靠和有效使用很关键。
The alignment, evaluation, and hallucination are three key areas of focus for ensuring the safe, reliable, and efficient use of LLMs. 
% 对齐（Alignment）：对齐是指确保模型的输出符合人类价值观和意图的过程。这包括确保模型不会产生有害的内容，如虚假信息、仇恨言论或偏见。对齐通常通过精心设计的训练数据、模型架构和人类反馈来实现。
Alignment refers to the process of ensuring that a model's outputs are consistent with human values and intentions. This involves preventing the generation of harmful content, such as disinformation, hate speech, or bias. Alignment is typically achieved through the use of carefully curated training data, the model's architecture, and incorporating human feedback. 
% 评估（Evaluation）：评估是指对模型的性能进行测试和分析，以确保其满足特定的标准和目标。这包括对模型的准确率、可解释性、公平性、安全性和其他指标进行评估。评估通常涉及多种方法和工具，包括人工审核、自动化测试和用户反馈。
Evaluation is the process of assessing and measuring the performance of a model to ensure that it meets certain standards and objectives. This includes evaluating the model for accuracy, interpretability, fairness, safety, and other metrics. Assessments typically involve a variety of methods and tools, such as manual review, automated testing, and user feedback. 
% 幻觉性（Hallucinations）：幻觉性是指模型生成的不准确或前后不一致的输出。这可能是由于模型对数据的错误理解或生成过程的随机性造成的。幻觉性是一个重要的挑战，特别是在需要高度准确性的应用中。因此，研究者们致力于开发减少幻觉性的方法，同时保持模型的创造力和灵活性。
Hallucination refers to outputs from a model that are inaccurate or inconsistent. These issues can arise from the model's misinterpretation of the data or the inherently stochastic nature of the generation process. The phenomenon of hallucination is a significant challenge, particularly in applications that demand a high level of precision. Consequently, researchers are actively working to develop strategies to minimize hallucination while preserving the creativity and adaptability of the model.

\subsection{AI Agents}
% 3D大模型
% 3D数据生成技术是一种利用大型模型生成3D模型的方法。这项技术通常需要使用大量的2D或3D数据来训练模型，使其能够理解和模仿3D物体的特征和结构。具体来说，它包括从多个视角生成图片，将这些图片合并，或者生成代码以调用3D建模软件来创建3D模型。通过这种方式，模型可以生成新的3D数据，这些数据在视觉上与现实世界中的物体相似，并且可以应用于各种领域，如游戏开发、虚拟现实、增强现实和3D打印等。随着技术的不断进步，这些模型正变得越来越准确和高效，能够生成越来越复杂的3D场景和物体。
3D data generation technology is a process that employs large models to produce 3D objects. This technology typically necessitates the use of extensive 2D or 3D data to train the model, enabling it to understand and replicate the characteristics and structure of 3D objects. Specifically, it involves generating images from multiple viewpoints, performing 3D reconstructions from these images, or generating code to call 3D modeling software to build 3D objects. In this manner, the model can generate new 3D data that resembles real-world objects visually and can be applied in various sectors such as game development, virtual reality, augmented reality, and 3D printing. As technology continues to advance, these models are becoming increasingly accurate and efficient, capable of generating more complex 3D scenes and objects.

% XoT
% XoT技术是一种用于理解和生成复杂文本以及进行逻辑推理的人工智能技术，例如“链式思考”和“程序式思考”等。通过让AI模型逐步解释和阐述其思考过程，XoT技术提高了理解和生成复杂文本的能力。这意味着AI模型在回答问题时，不仅提供最终答案，还能展示其推理步骤，就像人类一样。这样的做法有助于确保模型的回答更加准确和可解释，同时也为用户提供了更多的信息，以便进行进一步的分析和讨论。XoT技术对于需要复杂逻辑和推理能力的应用，如自然语言理解、机器翻译和智能助手等，具有重要意义。
XoT technology is designed to comprehend and generate complex text, as well as perform logical reasoning tasks such as "chain-of-thought" and "program-of-thought". By requiring LLMs to explain and elaborate on their thought processes in a step-by-step manner, XoT technology enhances the capability to understand and generate complex text. This approach ensures that when an LLM provides an answer to a question, it not only delivers the final response but also demonstrates the steps it took to reach that answer, much like a human would. This method is beneficial as it guarantees more accurate and understandable answers, and it also furnishes users with additional information for further examination and discussion. XoT technology is particularly relevant for applications that necessitate complex logical and reasoning capabilities, including natural language understanding, machine translation, and intelligent assistants.

% RAG
% 检索增强生成技术（RAG）是一种结合了检索和生成能力的人工智能技术。在生成文本、回答问题或进行对话时，RAG模型不仅可以根据其训练时的知识进行生成，还可以利用外部知识库进行检索，从而提供更准确、更相关的信息。这种技术能够提高模型的灵活性和实用性，使得模型在处理复杂任务时能够提供更全面、更准确、更实时的答案。
Retrieval-augmented generation (RAG) combines the LLM capabilities of retrieval and generation. When generating text, answering questions, or engaging in conversations, the RAG model can draw not only on the knowledge it was trained on but also retrieve information from external knowledge bases, providing more accurate and relevant responses. This technology enhances the flexibility and practicality of the model, enabling it to offer more comprehensive, accurate, and timely answers when handling complex tasks.

% AI Agents
% AI智能代理技术致力于解决如何执行复杂任务和回答问题。大型模型AI智能代理能够模拟人类的认知能力，具备个人资料、记忆、规划、行动的能力。通过调用各种工具，大型模型AI智能代理正变得越来越智能化，接近通用人工智能，能够执行越来越复杂的任务，并逐渐减少对人类干预的需求。
AI agents are dedicated to solving complex tasks and answering questions. LLM-based AI agents can simulate human cognitive abilities and possess capabilities such as profile, memory, planning, and action. By leveraging various tools, these agents are becoming increasingly intelligent, approaching general artificial intelligence (GAI). They can perform increasingly complex tasks and gradually reduce the need for human intervention.

% Long-context LLM
% 长上下文技术允许模型在训练和推理过程中考虑更长的上下文信息。这种技术对于自然语言处理的任务特别重要，它增强了模型的“记忆力”，使得模型能够处理更复杂的任务并表现出更强的通用性。通过采用长度外推和稀疏注意力等技术，解决了计算效率方面的问题。
Long context technology enables models to take into account longer context information during both training and inference. This technology improves the "memory" of the model, enabling it to tackle more complex tasks and exhibit greater versatility. Computational efficiency is addressed through the use of techniques like length extrapolation and sparse attention.

\subsection{AI PC}
% 主要基于云计算进行大语言模型的托管，他们没有将模型部署到AI PC。最近的工作开始将大语言模型部署到端侧，进行模型压缩。然而他们没有解决AIPC的网络安全问题。英伟达通过数字孪生，建设虚拟的数据中心仿真模拟，从而实现了真实策略的落地。
% 本文来介绍如何在共享环境中实现网络安全。
% 大模型端侧部署旨在将大型模型部署在个人计算机上，以增强个人计算机的性能和功能。此前，大模型主要部署在云计算中，这为低延迟和大规模普及带来了挑战。通过模型压缩和基于CPU、NPU、GPU的推理加速技术等，现在可以在个人计算机上直接运行大型语言模型（LLM）来处理复杂的AI任务，实现本地化部署，并且有助于保护用户数据隐私。AI PC的发展使得个人用户能够轻松地使用先进的AI功能，从而提高工作效率和生活质量。
AI PC is committed to promoting the deployment of LLM on personal computers to enhance the performance, functionality, and personalization of personal computers. Previously, LLMs were mainly deployed in cloud computing, which brought challenges to low latency and large-scale use. Through model inference acceleration technology based on CPU, NPU, and GPU, as well as model compression and quantization technology, LLMs can be run directly on personal computers to handle complex AI tasks, achieve localized deployment, and help protect user data privacy. The development of AI PCs enables individual users to easily use advanced and personalized AI functions to improve work efficiency and quality of life. 

\section{Approach}
\subsection{Automatic Evaluation of User Input Using Large Language Models}
\begin{figure}[!h]
\centering
\includegraphics[width=0.9\linewidth]{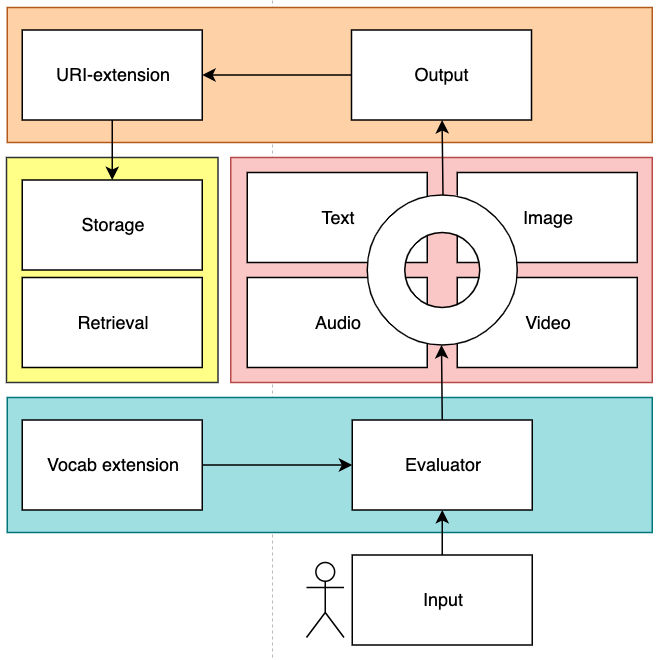}
\caption{Overview of the modular composition of the framework}
\label{arch.fig}
\end{figure}
% 图1显示了框架，其中红色、黄色、橙色区域代表meta 2.0中提到的可扩展性的多模态信息流，用于实现用户创意可控和责任可控。蓝色区域是本文中更新的输入评估模块，用于用户输入的评估。
The framework shown in Figure 1 outlines the use of scalable multi-modal information flow, as described in MetaAID 2.0 framework \cite{zhu2023metaaid}, to facilitate controllable user creativity and responsibility. The blue area represents the updated input evaluation module, which is integrated into the evaluation of user input. 
% 在维护绿色安全的网络环境方面，我们提出了一种对用户输入进行评估的方法。我们提出了五个评估指标。
To ensure ethical user interaction, we present a method to evaluate user inputs across the following 5 dimensions using LLMs.

% 道德准则：评估用户输入是否符合普遍接受的道德准则和伦理原则。例如，是否涉及歧视、仇恨言论、侵犯隐私等。
% 法律合规性：检查用户输入是否违反法律法规，包括但不限于版权、隐私法、反欺诈法等。
% 透明度：评估用户输入是否清晰、透明，是否有意误导或欺骗。
% 意图分析：分析用户输入的意图，判断是否隐藏不良动机或企图利用大语言模型进行恶意活动。
% 社会影响：考虑用户输入可能对他人或社会整体产生的影响，包括正面和负面影响。
(1) Ethics: Ensure user input aligns with widely accepted moral and ethical principles. This includes evaluating for discriminatory, hate speech, or privacy invasion content, among others. (2) Legal Compliance: Verify that user input does not violate any laws or regulations, such as copyright, privacy laws, or anti-fraud laws, to name a few. (3) Transparency: Assess whether user input is clear and transparent and whether it is intentionally deceptive or misleading. (4) Intent Analysis: Analyze the user's intent to determine if there are any hidden malicious intentions or attempts to use LLMs for nefarious activities. (5) Social Impact: Consider the potential positive and negative impacts that user input may have on others or society at large.

% 我们根据模型给出的每个方面的评分根据以下公式决定是否进行处理。
\begin{align}
\alpha = \sum_i sgn(ReLU(v_i - \tau_i))
\label{score}
\end{align}
where $\tau_i$ denotes the threshold for the $i$-th metric. 
% 只要其中m项超过了阈值，我们就拒绝这个输入。$\alpha\geq m$ 
So long as a certain number of metrics surpass the threshold $\alpha$, the system will not consider this user input. 

% 元宇宙中的用户输入包含多种语言以及各种表情包，目前的许多大模型都无法满足，例如Llama2的词汇表仅包含少量的中文，缺乏日语、韩语等多种语言。过大的词汇表会增加训练难度和造成模型的参数冗余。从头训练多语言的模型是昂贵且耗时的，已有的采用LoRA训练的模型无法扩展词汇表。理想的情况是针对不同的用户群体定制适应性的个性化的词汇表。为了解决这个问题，我们提出了词表扩展训练，可以按需扩展词表，仅需少量数据的微调便可以对模型进行适应，充分保留了模型的语言能力。评估用户输入是个高频且功能单一的场景，为了减少对输入进行监控的额外计算量，我们提出利用小模型结合词汇表扩展实现个性化的输入检测。
User inputs in the Metaverse \cite{zhu2022metaaid} include multiple languages and various emoticons, which many current large models cannot meet. For example, the vocabulary of Llama2-70b only contains a small amount of Chinese and lacks Japanese, Korean, and other languages. An excessively large vocabulary will increase the difficulty of training and cause parameter redundancy in the model. Training multilingual models from scratch is expensive and time-consuming, and existing models adapted with LoRA \cite{hu2021lora} cannot expand their vocabulary. The ideal situation would be to have an adaptive and personalized vocabulary tailored to different user groups. To solve this problem, we proposed vocabulary expansion training, which can expand the vocabulary on demand and adapt the model with only a small amount of fine-tuning of data, fully retaining the language ability of the model. Evaluating user inputs is a high-frequency and single-function scenario. To reduce the additional calculation amount for monitoring input, we propose to use small-sized LLMs combined with vocabulary expansion to achieve personalized input detection.

The training objective function of the model is maximum log-likelihood as shown in equation \eqref{obj}.
\begin{align}
min L = -\sum \log p(x_t|x_{<t},\Theta_{embed}, \Theta_{project})
\label{obj}
\end{align}
% 其中可学习的参数为embed和project，其他部分保持不变，大大降低了模型训练的显存消耗。在实验中，仅需要消耗56G显存就可以将词表扩展为49000的情况下，微调Llama2-70b。参数全部微调消耗的显存为70GB，因此效率提升了50%。模型通过最大似然损失来训练。
where $\Theta_{embed}$ and $\Theta_{project}$ are learnable parameters, and other parts remain unchanged, which greatly reduces the memory consumption of model training. In the experiment, Llama2-70b was fine-tuned while expanding the vocabulary to 49,000 by only optimizing 1.2\% parameters, so the GPU memory cost is significantly reduced. 

\subsection{Cybersecurity Simulation: A Practical Approach to Preparing for Cyber Threats}
% 在提升用户的自我保护意识方面，我们提出采用大模型生成许多网络安全相关的问答，以及生成网络攻击的代码。这种方法旨在通过实际操作和知识传授相结合的方式，帮助用户更好地理解和应对网络安全问题。
To enhance users' self-protection awareness, we adopt LLs to build a comprehensive simulation system that contains massive theoretical knowledge and practical cyber attack code. This strategy aims to assist users in better understanding and addressing cybersecurity issues by combining practical exercises with knowledge dissemination. 
% 通过大模型生成网络安全相关的问答，我们可以提供一个交互式的学习环境。用户可以在其中提出问题，并得到详细的解答。这样的问答内容可以包括如何识别钓鱼邮件、如何保护个人隐私、如何避免恶意软件的感染等。通过这种方式，用户可以在实践中学习，增强他们的网络安全意识。
Autoregressive LLMs generate text by predicting the next token, as shown in equation \eqref{gen}. By generating cybersecurity-related Q\&A using LLMs, we can create an interactive learning environment. Users can ask questions and receive responses on various topics. This approach allows users to enhance their cybersecurity awareness in the Metaverse. 
% 大语言模型通过next token prediction的形式生成文本。
\begin{align}
p(x) = \prod_{i=1}^n p(w_i|w_{i-1},...,w_1,\Theta)
\label{gen}
\end{align}
where $x$ is the input text and $w_i$ represents the $i$-th token. $\Theta$ is the model parameter.

% 我们采用了开源的大语言模型生成了元宇宙中的网络安全相关的问题，并且构建了一个包含1000个问题和答案对的数据集。然后我们采用对问题和答案随机采样的方式构建了一个考题生成的程序。
We used LLMs to generate questions related to cybersecurity in the Metaverse \cite{zhu2022metaonce}. We then compiled a dataset of 1,000 Q\&A pairs. Then we built a test question generation program by randomly sampling questions and answers. The user selects the correct option from multiple options, as shown in equation, as shown in equation \eqref{classifier}. The system also records wrong questions for users and provides corresponding suggestions.
\begin{align}
y = F_{classifier}(x^{(q)}_i, x^{(a)}_j)\qquad x^{(a)}_j \in X^{(a)}
\label{classifier}
\end{align}
% 这个游戏的规则是用户从多个选项中选出正确的选项。为用户记录错误的题目，并且给出相应的建议。
where $F_{classifier}()$ represents a user acting as a classifier, evaluating whether each response matches the corresponding question. $x^{(q)}_i$ and $x^{(a)}_j$ denote the $i$-th question and the $j$-th answer respectively. The set $X^{(a)}$ contains all the answers.

\begin{figure}[!h]
\centering
\includegraphics[width=0.7\linewidth]{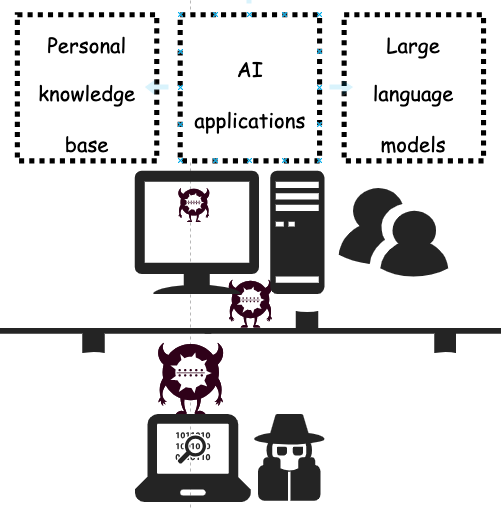}
\caption{Schematic diagram of a cyber attack on a localized deployment of LLM-based applications}
\label{cyber}
\end{figure}

% 生成网络攻击的代码，并且模拟攻击过程，可以帮助用户了解攻击者的手段和技术。在这个过程中，这些攻击的源代码并没有开放，而是仅提供了一个按钮。基于大语言模型辅助用户通过分析这些代码，了解攻击者如何利用系统漏洞进行入侵，从而更好地理解如何保护自己的网络安全。这种方式不仅可以提高用户的自我保护意识，还可以激发他们对网络安全技术的兴趣，鼓励他们进一步学习和探索。
Generating code for cyberattacks can help users understand attackers' methods and techniques. By using LLMs, users can analyze these codes to understand how attackers exploit system vulnerabilities \cite{das2022towards} to intrude, which in turn can enhance users' awareness of self-protection. This approach can also stimulate users' interest in cybersecurity technology and encourage them to further learn and explore, as shown in Figure \ref{cyber}.

% 我们分别采用了对话模型和代码生成模型生成攻击代码，例如codellama等。其中对话模型主要用于设计恶意提示模板，从而获取隐私信息。而代码生成模型用于生成恶意代码并且用于XSS攻击，并且解释代码的工作原理，然后用于攻击我们搭建的测试模型，并返回结果。用户模拟用户真正遭受攻击时，遭受信息泄漏的不同场景。
We used dialogue LLMs and code-generation LLMs to generate cyber attack prompts and code. The dialogue LLMs were primarily used to design malicious prompt templates to extract private information from localized deployment models. The code generation model was used to create malicious code for XSS attacks and to explain how the code works. We then used the code to attack the deployed target model and observed the results. Users simulated various scenarios where users experience information leakage during a real-world attack as below.
\begin{align}
response = F_{attack}(\psi(x), M)
\label{attack}
\end{align}
% 其中\psi是个包装，x是攻击的代码，M是被攻击大模型。这里攻击的形式是可选的，通过这种切身的体会，让用户观察大模型的输出。
where the blackbox wrapper $\psi()$ is designed to conceal the attack source code $x$. The attack code $x$ is utilized to evaluate the robustness of the target model $M$. The attack action $F_{attack}()$ encompasses various strategies, enabling users to observe the output of the target model.

% 我们还提供了用户反馈机制，允许用户对生成的内容进行评价和反馈。这样可以帮助我们不断优化生成的内容，提高其质量和相关性，更好地满足用户的学习需求。提升用户的自我保护意识是一个长期的过程。我们需要不断地提供新的内容和更新，以适应不断变化的网络安全威胁和用户需求。通过持续的培训和教育，我们可以帮助用户建立强大的网络安全防线。
We also provide a user feedback mechanism that allows users to evaluate and provide feedback on the generated content. This can help us continuously optimize the content we generate, improve its quality and relevance, and better meet the learning needs of users. Enhancing users’ self-protection awareness is a long-term process. We need to continuously provide new content and updates to adapt to evolving cybersecurity threats and user needs. Through ongoing training and education, we can help users build a strong cybersecurity defense.

\section{Experiments}
\subsection{Setup}
We build online demos to support user interaction and test model results. These experiments are run on an Intel(R) Xeon(R) Platinum 8163 CPU @ 2.50GHz with 256G of memory, and 8 RTX 3090 (24G) GPUs. 

\subsection{Results of Input Evaluation}
% 模型在纯文本输入的5个维度的评估上已经取得了较好的效果。此处我们选取了一个更具挑战性的图像输入的示例，如图1所示的一张广告图片，来实验多模态大语言模型在五个维度上对用户输入评估能力。
The model has already demonstrated good performance in evaluating plain text input across 5 dimensions. To further test the capabilities of the multimodal LLMs, we use a more challenging example of image input to assess its ability to process user input in these 5 dimensions. 

% 实验结果如表1所示，我们选取了几个代表性的视觉问答模型。其中--代表模型仅输出了描述性的文本段落但是没有输出评分。可以看出BLIP2、MiniGPT-v2能够一定程度上完成这个任务，但是他们的结果存在随机性。模型在回复生成用户期望输出的性能还有待提升。
The experimental results are shown in Table \ref{res}. We selected several representative visual QA models. Among them, scores denoted by - meaning that they only output descriptive text paragraphs but do not output scores. It can be seen that BLIP2 and MiniGPT-v2 perform better in completing this task. 
\begin{table*}[htbp!]
\centering
\caption{Image input evaluation scores (0-10) across 5 dimensions}
% \resizebox{\linewidth}{!}{
\begin{tabular}{c|ccccc} \toprule
Models   & Ethics &  Legal Compliance   &  Transparency &   Intent Analysis & Social Impact \\ \midrule
ViLT \cite{kim2021vilt} &  -- &  -- & --  & -- & -- \\
LLaVA \cite{liu2023visual} & --  &  -- & --  &  -- &  -- \\
BLIP-2 \cite{li2023blip} & 7  & 5  & 5  &  4 & 6 \\
MiniGPT-4 \cite{zhu2023minigpt} & 4 &  4 & --  &  -- & -- \\
MiniGPT-v2 \cite{chen2023minigpt} & 7  & 7  & 9 & 7 & 5 \\
\bottomrule       
\end{tabular}
% }
\label{res}
\end{table*}
\begin{figure}[!h]
\centering
\includegraphics[width=0.7\linewidth]{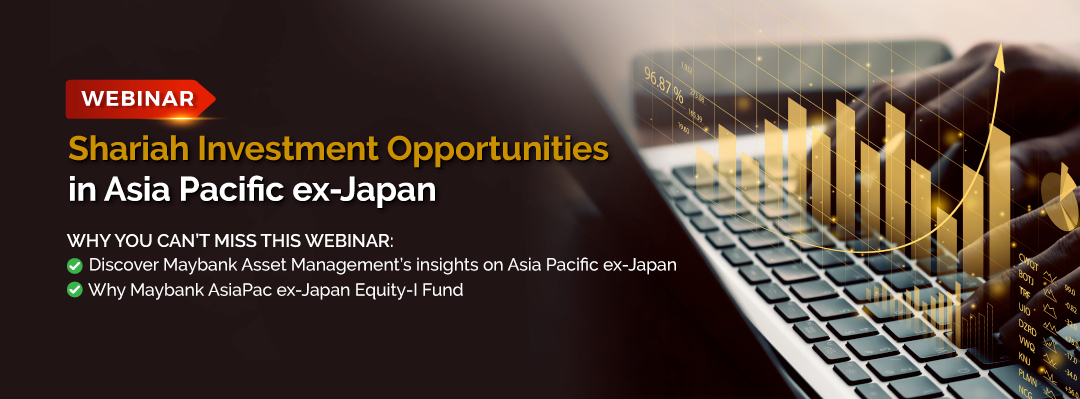}
\caption{An example of an input image about financial service}
\label{arch.fig}
\end{figure}

\subsection{Results of Cybersecurity QA}
% 我们采用一个示例来比较不同大模型对于用户元宇宙安全教育方面的效果。可以看出虽然模型都能产生对网络安全有帮助的回复，但是大部分都针对传统的网络安全，仅有少部分的回复是针对元宇宙安全的。因此在元宇宙网络安全知识注入还需进一步提升。
We use an example to compare how different LLMs respond to user questions about Metaverse cybersecurity. \texttt{How can users take actions to prevent and reduce virtual social risks they may face in the Metaverse, such as encountering fraud, harassment, privacy violations, etc.?} The example illustrates that while various LLMs can generate responses that are helpful for cybersecurity, most of these responses focus on traditional cybersecurity concerns, as shown in Table \ref{comparison1} and \ref{comparison2}. Only a limited number of responses address the unique security challenges of the Metaverse. This suggests that there is room for improvement in integrating cybersecurity knowledge into the Metaverse context.
% 我们以示例的形式给出安全问答模拟的结果。
\begin{CJK*}{UTF8}{gbsn}
\begin{table*}[]
\centering
\tiny
\caption{Comparison of responses from different models (part 1)}
% \resizebox{\linewidth}{!}{
\begin{tabularx}{\linewidth}{c|X } \toprule
Models   & \multicolumn{1}{c}{Response}
\\ \midrule 
Phi-2-2.7B \cite{li2023textbooks} & To take actions to prevent and reduce virtual social risks in the Metaverse, users can apply the following principles:

- Follow the rules and etiquette of the game or platform they are using, such as agreeing to terms and conditions, respecting others' identities and preferences, and adhering to the community standards and codes of conduct.

- Check the reliability and security of the servers, devices, and networks they are using, and use antivirus and firewall software, strong passwords, and encryption methods to protect their personal information and data.

- Monitor and report any suspicious or malicious behavior or activities they encounter or witness, such as phishing emails, fake profiles, scams, hate speech, cyberbullying, etc., and seek help from the support teams or authorities if needed.

- Limit the amount of time and money they spend on the game or platform, and balance their online and offline activities and responsibilities, and prioritize their physical and mental health and well-being.\\ \midrule
ChatGLM3-6B \cite{zeng2022glm}&To prevent and reduce virtual social risks in the Metaverse, users can take the following actions:

Research and choose a reputable Metaverse platform: Before joining a Metaverse platform, users should research and choose a reputable platform that has a good reputation and effective measures to protect users' rights and interests.

Use secure communication tools: Users should use secure communication tools, such as encrypted messaging apps and video conferencing software, to protect their personal information and prevent virtual attacks.

Learn about virtual currencies and economic systems: Users should learn about virtual currencies and economic systems in the Metaverse to avoid being defrauded or scammed by unscrupulous individuals or organizations.

Be cautious with personal information: Users should be cautious with their personal information, such as their real name, address, and financial information, and should not share them with strangers or untrusted sources.

Report suspicious activities: If users encounter any suspicious activities, such as fraud, harassment, or privacy violations, they should report them to the relevant authorities or Metaverse platforms immediately.

Install anti-virus software: Users should install anti-virus software to protect their devices from malware and other virtual threats.

Use privacy-preserving tools: Users can use privacy-preserving tools, such as VPNs and Tor, to protect their online identity and activities from being monitored or accessed by others.

Educate themselves and others: Users should educate themselves and others about the risks and challenges of the Metaverse and how to avoid them to promote a safer and more secure virtual environment.

By taking these measures, users can reduce the risks of encountering fraud, harassment, privacy violations, and other social risks in the Metaverse. \\ \midrule
Mistral-7B \cite{jiang2023mistral}& There are several actions that users can take to prevent and reduce virtual social risks in the Metaverse. Here are some suggestions:

Use strong and unique passwords: It is important to use strong and unique passwords for all your Metaverse accounts. Avoid using the same password for multiple accounts and use a combination of letters, numbers, and symbols.

Enable two-factor authentication: Enable two-factor authentication on your Metaverse accounts to add an extra layer of security. This will require you to enter a code sent to your phone or email before you can access your account.

Be cautious of phishing scams: Be cautious of phishing scams, which are attempts to trick you into revealing personal information or clicking on malicious links. Always verify the sender's identity and check the link before clicking on it.

Report any suspicious activity: If you encounter any suspicious activity, such as fraud, harassment, or privacy violations, report it immediately to the Metaverse platform or the relevant authorities.

Use privacy settings: Use privacy settings on your Metaverse accounts to control who can see your personal information and how it is shared.

Be mindful of your online behavior: Be mindful of your online behavior and avoid sharing personal information or engaging in activities that could put you at risk.

Use reputable sources: Use reputable sources when making purchases or interacting with other users in the Metaverse. Avoid clicking on links or downloading attachments from unknown sources.

By following these actions, users can reduce the risk of encountering virtual social risks in the Metaverse.\\ \midrule
Zephyr-7B \cite{tunstall2023zephyr}& Use reputable platforms: Users should only use Metaverse platforms that have a good reputation and are known for their security measures. They should research the platform's privacy policies, data protection practices, and user feedback before creating an account.

Create strong passwords: Users should create unique and strong passwords for their Metaverse accounts and change them regularly. They should also enable two-factor authentication to add an extra layer of security.

Limit personal information: Users should avoid sharing sensitive personal information, such as their real name, address, or phone number, in the Metaverse. They should also be cautious about sharing their location or other identifying details.

Be aware of scams: Users should be wary of unsolicited messages, offers, or requests for money or personal information. They should verify the authenticity of any offers or requests before responding.

Report harassment: If users encounter harassment or abuse in the Metaverse, they should report it to the platform's moderators or administrators. They should also document the incident and save any evidence to support their report.

Use privacy settings: Users should adjust their privacy settings to control who can see their content and interact with them. They should also limit the amount of personal information they share publicly.

Educate oneself: Users should stay informed about the latest Metaverse trends, risks, and best practices. They should also attend online safety workshops and seminars to learn how to protect themselves in the Metaverse.

Use common sense: Users should use common sense and trust their instincts when interacting with others in the Metaverse. They should avoid meeting strangers in person, sharing passwords, or engaging in risky behavior.

Limit time spent: Users should limit the amount of time they spend in the Metaverse to avoid becoming addicted or isolated from the real world. They should also take breaks and disconnect from the Metaverse to recharge and refocus.

Seek help: If users experience any mental health or well-being issues related to their Metaverse use, they should seek professional help or support from trusted sources. They should also prioritize their physical and mental health over their virtual experiences.\\ \midrule
OpenChat-7B \cite{wang2023openchat}& Users can take several actions to prevent and reduce virtual social risks in the Metaverse, such as encountering fraud, harassment, privacy violations, and more. Here are some steps users can take:

Set strong passwords: Use unique and complex passwords for your Metaverse accounts to prevent unauthorized access. Enable two-factor authentication (2FA) whenever possible for added security.

Be cautious with personal information: Avoid sharing sensitive information like your full name, address, phone number, or financial details with strangers or suspicious entities.

Use a reputable virtual environment: Choose a well-established and secure virtual environment with a good track record of protecting user data and privacy.

Familiarize yourself with the platform's privacy settings: Understand and configure the privacy settings of the platform you are using to control who can access your information and interactions.

Report suspicious behavior: If you encounter any suspicious or malicious activity, report it to the platform's administrators or moderators.

Be cautious with third-party apps and services: Only use trusted third-party apps and services, and ensure they have a solid reputation for security and privacy.

Be cautious with in-platform transactions: When engaging in transactions within the Metaverse, use trusted methods and platforms. Avoid sharing financial information with unverified entities.

Use a VPN: A Virtual Private Network (VPN) can help protect your privacy by encrypting your internet connection and masking your IP address.

Educate yourself: Stay informed about the latest threats and best practices for staying safe in the Metaverse. Regularly update your knowledge of security measures and privacy settings.

Be cautious with social engineering attacks: Be wary of phishing emails, scams, and other attempts to trick you into revealing personal information or downloading malicious software.

By taking these steps, users can significantly reduce the risks they may face in the Metaverse and enjoy a safer and more secure virtual experience.\\ \midrule
Baichuan2-13B \cite{yang2023baichuan}& In order to prevent and reduce virtual social risks in the meta-universe, users should take the following measures: 

1. Understand the rules and regulations of the platform: Users need to understand the relevant regulations and risk prevention knowledge of the platform before using it so that they can better protect their rights and interests.

2. Protect personal information: Do not easily disclose personal sensitive information on the Internet, including but not limited to names, addresses, phone numbers, bank account information, etc.

3. Use security software: Install anti-virus software and firewalls to protect your computer from malicious attacks and viruses.

4. Avoid random connection with strangers: In the virtual world, you should avoid connecting with strangers at will to avoid being deceived or harassed.

5. Check the source of information: Before believing or spreading the information released by others, make sure its authenticity through third-party verification tools.

6. Pay attention to online behavior: Do not perform illegal behaviors, such as hacking, phishing, etc., which may lead to legal problems.

7. Educate yourself and others: By reading related books and articles, understanding the latest developments and risks in the meta-universe, and sharing these insights with other people to help them better understand and deal with potential risks.\\ \midrule
Yi-34B \cite{Yi2023}& Users can take several actions to prevent and reduce virtual social risks in the Metaverse:

Educate Yourself: Stay informed about common scams, fraud tactics, and privacy risks. Understanding how virtual environments work and potential threats can help you make better decisions and spot potential dangers.

Choose Wisely: Select reputable platforms and applications with a strong track record of user safety and privacy. Research the reputation of Metaverse platforms before using them.

Manage Your Digital Footprint: Be mindful of the personal information you share in the Metaverse. Limit the amount of sensitive data you disclose, and consider using pseudonyms or avatars to protect your identity.

Secure Your Accounts: Use strong, unique passwords for your Metaverse accounts. Enable two-factor authentication (2FA) whenever possible to add an extra layer of security.

Be Cautious with Transactions: If you engage in virtual transactions, use secure payment methods and be wary of offers that seem too good to be true. Only deal with reputable parties and be cautious of requests for personal information during transactions.

Control Your Privacy Settings: Familiarize yourself with the privacy settings available in the Metaverse platforms you use. Adjust these settings to limit who can contact you, view your content, or access your data.

Report and Block: If you encounter harassment, fraud, or privacy violations, report the incident to the relevant authorities and the platform's moderation team. Block and avoid interaction with users who engage in inappropriate behavior.

Stay Updated: Regularly update the software and apps you use in the Metaverse to ensure you have the latest security features and bug fixes.

Seek Support: If you encounter issues, seek support from the platform's customer service or community forums. They may be able to provide guidance on how to handle specific situations.

Be Smart About Interaction: Avoid disclosing personal information, getting involved in controversial topics, or engaging in activities that could lead to negative interactions with others.

Remember, while these precautions can help reduce risks, no system is foolproof. Always be vigilant and cautious when navigating the Metaverse.\\

\bottomrule       
\end{tabularx}
% }
\label{comparison1}
\end{table*}
\end{CJK*}

\begin{CJK*}{UTF8}{gbsn}
\begin{table*}[]
\centering
\tiny
\caption{Comparison of responses from different models (part 2)}
% \resizebox{\linewidth}{!}{
\begin{tabularx}{\linewidth}{c|X } \toprule
Models   & \multicolumn{1}{c}{Response}
\\ \midrule 
Mixtral-8x7B & Users can take several steps to prevent and reduce virtual social risks in the Metaverse:

1. Educate yourself about online safety: Before entering any virtual world, educate yourself about potential risks and learn how to protect your personal information and assets. Familiarize yourself with security measures provided by the platform and keep yourself updated on new threats and scams.

2. Use strong passwords and two-factor authentication (2FA): Create complex passwords that include a mix of letters, numbers, and special characters. Enable 2FA whenever possible for an extra layer of protection. Regularly change your passwords and avoid sharing them with others.

3. Limit personal information shared online: Be cautious when disclosing personal details like your real name, address, phone number, or financial information. Configure your account settings to limit visibility of this data and use pseudonyms if appropriate.

4. Customize privacy settings: Make sure you understand and configure privacy settings within the Metaverse platform. This will help control who can access your content, interact with you, or view your personal information.

5. Verify identities before transactions: When buying or selling items, ensure you are dealing with reputable individuals or organizations. Double-check their identity using multiple sources and never share sensitive information during initial interactions.

6. Report suspicious behavior: If you come across fraudulent activities, harassment, or other forms of abuse, report it immediately to the platform's moderators or support team. Provide detailed information and evidence to assist in investigating the issue.

7. Block unwanted contacts: Most platforms allow you to block users who engage in inappropriate behavior or make you uncomfortable. Utilize these features to maintain a safe environment.

8. Join trusted communities: Participate in groups, guilds, or communities vetted for positivity and respectful engagement. Avoid joining unknown groups where members might not adhere to community guidelines.

9. Stay vigilant: Keep an eye out for phishing attempts, malware, and other scams. Do not click on suspicious links or download unfamiliar files from untrusted sources.

10. Encourage positive culture: Help create a supportive atmosphere within the Metaverse by promoting kindness, empathy, and respect towards others. Stand up against negative behaviors and encourage responsible usage among fellow users.

11. Establish boundaries: Clearly communicate your expectations regarding interaction and set limits for what is acceptable behavior. Don't hesitate to remove yourself from situations that feel unsafe or uncomfortable.

12. Parental controls: If children are involved, enable parental controls and monitor their activity within the Metaverse. Teach them about online safety and supervise their interactions with others.\\ \midrule
Llama2-70B \cite{touvron2023llama} &Great question! There are several steps that users can take to prevent and reduce virtual social risks in the Metaverse. Here are some suggestions:

1. Use strong passwords and two-factor authentication: Users should create strong passwords and enable two-factor authentication to protect their accounts from hacking and unauthorized access.

2. Be cautious of suspicious behavior: Users should be aware of their surroundings and be cautious of suspicious behavior, such as unsolicited messages or friend requests from strangers. They should also be wary of clicking on suspicious links or downloading attachments from unknown sources.

3. Adjust privacy settings: Users should adjust their privacy settings to control who can see their virtual environment, items, and personal information. They should also be mindful of what they share online and with whom they interact.

4. Use reputable virtual currencies: Users should only use reputable virtual currencies and avoid engaging in transactions with unverified parties. They should also be aware of the risks associated with virtual currency transactions, such as fraud and market volatility.

5. Use secure communication channels: Users should use secure communication channels, such as end-to-end encrypted messaging apps, to protect their conversations from eavesdropping and interception.

6. Stay informed: Users should stay informed about the latest virtual social risks and trends, as well as the measures they can take to protect themselves. They should also keep their software and applications up to date to ensure they have the latest security patches.

7. Use virtual reality safety features: Users should use virtual reality safety features, such as limiting their field of view, to prevent unauthorized access to their virtual environment. They should also be aware of their physical surroundings while using virtual reality and take regular breaks to avoid eye strain and other health issues.

8. Engage in safe virtual interactions: Users should engage in safe virtual interactions by avoiding controversial or sensitive topics, not sharing personal information, and being respectful of others' privacy and boundaries. They should also be aware of cultural differences and avoid engaging in behavior that may be considered offensive or inappropriate.

9. Use trusted virtual reality platforms: Users should use trusted virtual reality platforms that have a good reputation for security and privacy. They should also read reviews and do research before selecting a virtual reality platform to ensure it meets their security and privacy needs.

10. Seek help when needed: Users should seek help when needed, such as reporting suspicious behavior or seeking assistance from virtual reality platform customer support. They should also be aware of their own mental health and take breaks from virtual reality when needed to avoid addiction or other negative effects.

By following these steps, users can reduce their exposure to virtual social risks and enhance their overall safety and privacy in the Metaverse. \\ \midrule
% Goliath-120B & \\ \midrule
Falcon-180B \cite{almazrouei2023falcon}& There are several ways that users can protect themselves from virtual social risks when using the Metaverse. Some of these include:

1. Use strong passwords and enable two-factor authentication (2FA) whenever possible.

2. Be cautious about who you interact with online and avoid sharing personal information with strangers.

3. Be aware of potential phishing scams or fake websites designed to steal your personal information or login credentials.

4. Keep your devices up to date with the latest security patches and updates.

5. Install reputable anti-virus software on your device and regularly scan it for malware.

6. Verify the authenticity of any virtual goods or services before making a purchase or downloading them.

7. Always use trusted platforms or applications when communicating with others in the Metaverse.

8. Report any suspicious activity or incidents immediately to the platform or authorities if necessary.

9. Be mindful of privacy settings and adjust them accordingly to control what information is visible to others. By following these best practices, users can significantly reduce their risk of experiencing virtual social issues in the Metaverse.\\ 
\bottomrule       
\end{tabularx}
% }
\label{comparison2}
\end{table*}
\end{CJK*}

\section{Conclusion}
% 本文旨在通过虚拟仿真技术来提高元宇宙中的大模型安全意识。具体的我们提出利用大语言模型生成问答测试和恶意代码，以进行模拟和演练，从而提升用户抵御风险的能力。为了解决元宇宙中的用户输入内容的安全问题，我们提出了LLM评估方法解决用户输入安全问题，提出词汇扩展学习从而实现个性化用户群体语言需求，最终实现更快、更安全、更具创新性的元宇宙智能软件开发。
This paper aims to enhance cybersecurity awareness in the Metaverse by utilizing virtual simulation technology based on LLMs. Specifically, we propose using LLMs to generate theoretical knowledge and potentially malicious code for simulation purposes, to improve users' ability to defend against cyber attacks within the Metaverse. To ensure ethical user interactions, we employ an LLM evaluation method that assesses across 5 dimensions. We also present a vocabulary expansion training method to adapt to personalized user groups, aiming to achieve faster, safer, and more innovative Metaverse intelligent software development.

% 本文的创新点在于提出了一种新颖的元宇宙应用的大模型安全机制。在未来，我们希望将该方法应用于本地化部署的大语言模型中，例如包含个性化信息的AI PC以及数字人应用。
The innovation of this paper lies in the use of LLMs to enhance the cybersecurity mechanisms for Metaverse applications. We aim to apply this method to locally deployed LLMs in the future, including scenarios like AI PC security and digital human applications that involve personalized information.

\bibliographystyle{IEEEtran}
\bibliography{reference}

\end{document}